\documentclass[twocolumn,amssymb,floatfix,aps,prb,showpacs,citeautoscript]{revtex4-1}

\usepackage[dvips]{graphicx}
\usepackage{graphicx}
\usepackage[utf8]{inputenc}
\usepackage{mathtools}
\usepackage{dcolumn}
\usepackage{physics}
\usepackage{bbold}
\usepackage{bm}
\usepackage{graphics}
\usepackage{dirtytalk}
\usepackage{epsfig,color}
\usepackage{xcolor,cancel}
\usepackage[breaklinks,colorlinks,bookmarks=false,citecolor=red,linkcolor=blue,urlcolor=magenta]{hyperref}
\graphicspath{{Figures/}} 

\begin{document}
\title[J. Vahedi]{Dynamical quantum phase transitions in  Stark quantum spin chains}

\author{M. Faridfar}
\affiliation{Department of Physics, Sari Branch, Islamic Azad University, Sari 48164-194, Iran.}

\author{A. Ahmadi Fouladi}
\affiliation{Department of Physics, Sari Branch, Islamic Azad University, Sari 48164-194, Iran.}

\author{J. Vahedi}
\email{j.vahedi@jacobs-university.de}
\affiliation{Department of Physics and Earth Sciences, Jacobs University Bremen, Bremen 28759, Germany.}
\affiliation{Department of Physics, Sari Branch, Islamic Azad University, Sari 48164-194, Iran.}

\date{\today}
\begin{abstract}
We investigate the nonequilibrium dynamics of one-dimension spin models in the presence of a uniform force. The linear potential induces delocalization-localization transition in the free particles model which is known as the Wannier-Stark effect. We study dynamical quantum phase transition (DQPT) due to sudden global quenches across a quantum critical point when the system undergoes a localization-delocalization transition. In this regard, we consider the XX and XXZ spin chains and explore two types of quenches with and without ramping through the delocalization-localization point. The XX model was mapped to the free fermion particles, so both analytical and numerical results were provided. Results unveil that the dynamical signature of localization-delocalization transition can be characterized by the nonanalyticities in dynamical free energy (corresponds to the zero points in the Loschmit echo). We also explore the interaction effects considering XXZ spin chains, using the time-dependent extension of the numerical DMRG technique. Our results show that depending on the anisotropic parameter $\Delta\lessgtr1.0$, if both the initial and post-quench Hamiltonian are in the same phase or not, DQPTs may happen. Moreover, the interrelation between DQPTs with different correlation measures such as the equilibrium order parameters or entanglement entropy production of the system remains unclear. We provide more analyses on the feature of DQPTs, in both types of quenches, by connecting them to the average local magnetization, entanglement entropy production, and the Schmidt gap.
\end{abstract}

\maketitle
\section{Introduction}\label{sec1}
Phase transitions are considered a fascinating phenomenon in physics as a small change in the system parameter can drive the model between different phases which are not connected adiabatically. This is accompanied by nonanalyticities in the free energy. A generalization of this fundamental concept to the nonequilibrium quantum realm has triggered the dynamical quantum phase transition (DQPT)\cite{Heyl2013}. To describe the dynamics of a quantum system that is perturbed by a sudden change of the Hamiltonian, an important quantity is the Loschmidt echo, which measures the overlap of the initial quantum state and the time evolved state after the quench\cite{Heyl2018}. The signature of a dynamical phase transition is captured by vanishing the Loschmidt echo in the thermodynamic limit. This corresponds to the nonanalyticities at a certain time in the dynamical free energy and is found to have a connection with the dynamics of order parameters\cite{Halimeh2021}. 
\par
DQPTs  studied intensively in many models such as transverse field Ising model\cite{Heyl2013}, anisotropic XY model\cite{Vajna2014}, quantum Potts chain\cite{Karrasch2017},  Hubbard and Falicov-Kimball models\cite{Canovi2014}, two-band topological systems\cite{Szabolcs2015,Huang2016,Budich2016,Wing2021}, flat band networks\cite{Hao2019},and bosonic models\cite{Dolgitzer2021,Syed2021}. DQPTs were also studied in inhomogeneous models such as the random energy model\cite{Obuchi2012,Takahashi2013}, the Anderson model\cite{Honghao2018}, as well as quasiperiodic potential in the context of the Aubry-Andre model\cite{Yang2017,Modak2021}. Exploring DQPTs in higher dimensions reported on models such as the integrable two-dimensional Kitaev honeycomb\cite{Schmitt2015}, two-dimensional Haldane\cite{Bhattacharya2017,Dutta2017}, three-dimensional O(N) \cite{Weidinger2017} models, and two-dimensional transverse field Ising model\cite{Markus2018,Halimeh2022}. Within the current state-of-art experimental techniques, namely trapped-ion and ultracold-atom, DQPTs are observed in the dynamics of transverse-field Ising models\cite{Jurcevic2017} and the dynamical topological quantum phase transitions\cite{Tarnowski2018}.
\par
Initially, it is argued that the occurrence of DQPTs need that the quench process is ramped through an equilibrium quantum critical point\cite{Heyl2013}. This has been explored in many studies for both non-integrable\cite{Canovi2014,Karrasch2014,Halimeh2017,Hubig2019} and integrable\cite{Karrasch2013,Schmitt2015,Dutta2017,Dutta2018,Ricardo2020}models. Further theoretical support also pointed out that for quenches not belonging to this class, the so-called "accidental" DQPTs can still occur, requiring a fine-tuning of the Hamiltonian\cite{Liu2019,Enss2012,Andraschko2014,Amina2019,Sedlmayr2020,Ding2020} . 
\par
Apart from quantum phase transitions governed by the Landau theory, the localization-delocalization transition also surged many interests in the field. The seminal idea was introduced by Anderson with the quenched disorder (randomness on the Hamiltonian of interest)\cite{Anderson1958}. A closely related setting for localizing single particles considers quasiperiodic potential in the context of the Aubry-Andr\'e (AA)\cite{Aubry1980}. Another well-known mechanism for localizing single particles is the Wannier-Stark effect\cite{Wannier1960}, in which particles living on a lattice become localized in the presence of a linear potential. Later, many-body localization born to answer the nature of quantum systems when both disorder and interactions are present. In the seminal perturbative calculations, Basko, et.al;\cite{Basko2006} have proved that many-body eigenstates can also localize\cite{Polyakov2005,Huse2010,Abanin2019}.
\par
In this paper, we focus on the possible DQPTs in the quantum spin chains treated with a tilted field. In this model, single particles are localized with the presence of the tilted field (known also as Bloch localization). The effects of interaction in the context of many-body localization has been recently introduced\cite{Refael2019} and explored\cite{Guo2021,Yao2021,Morong2021,Herviou2021,Li2021,Vernek2022,Zisling2022}. Quench dynamic in this model has been studied analytically for XX chain\cite{Antal1999,Peschel2009} and XXZ model\cite{Mitra2010}, but did not consider the model within the DQPT context, which is the aim of this work. What we are interested in is exploring DQPTs by quenching the model within different parameters space. This mainly is divided into two types of quenches with and without ramping through a quantum critical point. To this end, we consider the general $H_{\rm xyz}$ spin chains modulated with increasing linear field (with strength $\mathcal{F}$). Using both analytical and numerical techniques, we solve the model to find the eigenstates (or the ground) state before and after quench and probe the overlap between them over time. 
\par
The rest of the paper is organized as follows. In Sec. \ref{sec2}, we introduce the model and study the quench dynamics with the changing field strength between $\mathcal{F}\rightarrow0~(\mathcal{F}\rightarrow\infty)$ and $\mathcal{F}\rightarrow\infty~(\mathcal{F}\rightarrow0)$ . We give analytical (numerical) proof of DQPT for the non-interacting XX model ( interacting XXZ model ) and demonstrate that there are a series of zeros of Loschmidt echo, as indicates the DQPTs indeed occur. Conclusion are summarized in Sec.\ref{sec3}. In appendix~\ref{appA}, we address the numerical finite size effects in the XXZ model. In appendix~\ref{appB}, we present the derivation of local observables for free fermion when particles are exposed the linear potential.

\section{DQPT AND MODELS}\label{sec2}
To probe DQPTs, we consider the Loschmidt amplitude (return amplitude) which has been introduced in the seminal paper\cite{Heyl2013}
\begin{equation}
\mathcal{G}(t)=\bra{\Psi_i}\ket{\Psi_i(t)}=\bra{\Psi_i}e^{-iH_ft}\ket{\Psi_i}
\label{eq1}
\end{equation}
where $\ket{\Psi_i}$ denotes the quantum initial ( localized or the ground) state that pertains to Hamiltonian before quench and evolves under quenched Hamiltonian $H_f$ in time. The corresponding Loschmidt echo (LE) $\mathcal{L}(t)$ is given by the squared modulus of the Loschmidt amplitude $\mathcal{L}(t)=|\mathcal{G}(t)|^2$. It has been shown, that if the post-quench Hamiltonian $H_f$ and the initial Hamiltonian $H_i$ correspond to different phases, the LE exhibits a series of zeros at some critical times ${t^*}$ in the thermodynamic limit. Analogous to the equilibrium phase transition theory, a dynamical free energy density 
\begin{equation}
f(t)=-\lim_{N \to +\infty} \frac{1}{N} \ln\mathcal{L}(t)
\label{eq2}
\end{equation}
in the thermodynamic limit becomes nonanalytic as a function of time, which is regarded as a signature of the DQPT and tested in various systems~\cite{Heyl2013,Heyl2018,Bozhen2019,Mirmasoudi2019,Tong2021,Nicola2021}. 
\par
We consider XYZ  Hamiltonian in a tilted field
\begin{equation}
H_{\rm xyz} =-\sum_
n\Big[J_xS_n^xS_{n+1}^x+J_yS_n^yS_{n+1}^y+J_z S_n^zS_{n+1}^z-h_nS_n^z\Big]
\label{eq3}
\end{equation}
where $S_n^\nu~\{\nu=x,y,z\},$ are spin-$1/2$ operators on the $n$-th site, $h_n=\mathcal{F}n$ is a linearly varying magnetic field in the $z$ direction, $J_\nu$ are the exchange couplings. The ground state properties of spin chains in various spatially varying magnetic fields have been studied in the literature. 
We consider Eq.(\ref{eq3}) in two parameters limit as (i) non-intereacting XX model with $J_x=J_y$ and $J_z=0$, and (ii)  interacting XXZ model with $J_x=J_y\neq J_z$. 
 \begin{figure}[t]
 \includegraphics[width=1\columnwidth]{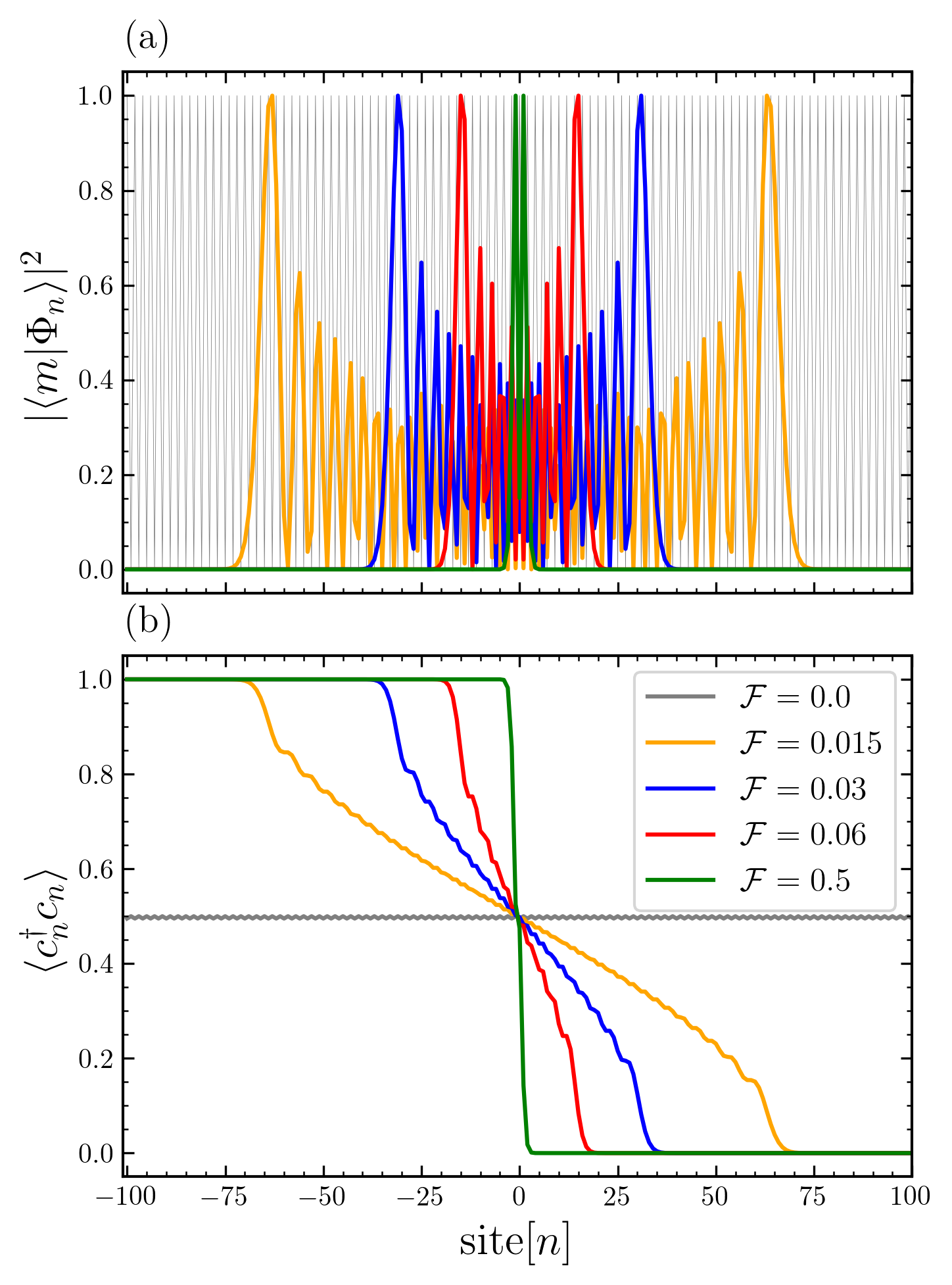}
 \caption{(colour online)  (a) Density profile $|\langle m|\Phi_n\rangle|^2$ as a function of the site index $i$ for a state in the middle of spectrum $\epsilon_m=0$. (b) The ground state particles occupation (or magnetization in spin language $m^z_n=\langle c_n^\dagger c_n\rangle-\frac{1}{2}$) profile of the model. Results of different field strengths are shown in colored solid lines. System size is $N=201$. }
\label{fig1}
\end{figure}
 \begin{figure}[t]
 \includegraphics[width=1\columnwidth]{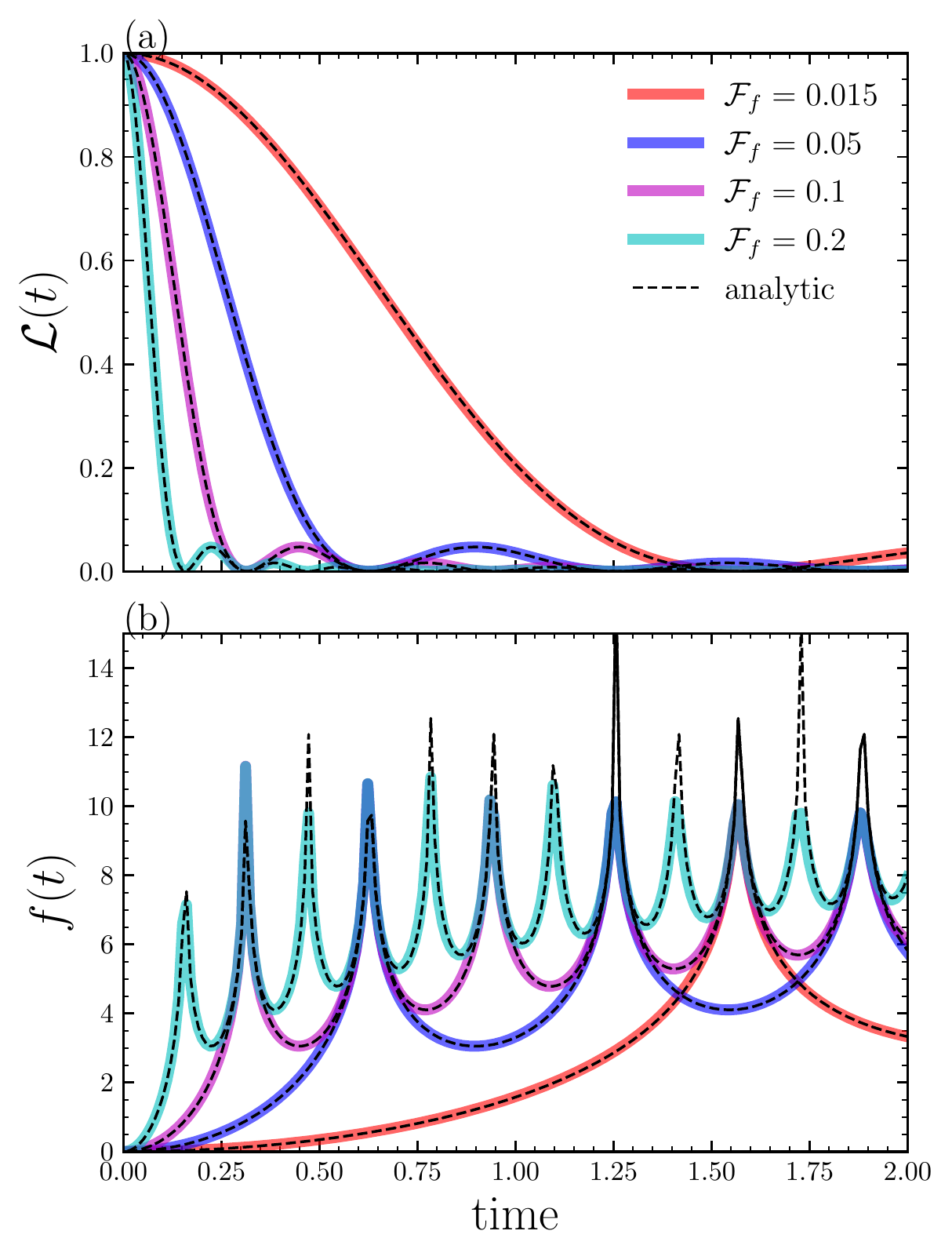}
 \caption{(colour online) Results for the forward quench from a plane wave state ($\mathcal{F}=0$) to a localized state ($\mathcal{F}\neq0$). (a) evolution of Loschmidt echo for the XX model with $\mathcal{F}_f=0.015,0.05,0.1$ and $\mathcal{F}_f=0.2$. (b) corresponding dynamical free energy. The black dashed curve corresponds to the analytical result $\mathcal{G}(t)=|\sin{x}/x|^2$. The initial state is fixed to be one of the plane wave states in the middle of the spectrum of the initial Hamiltonian with $\mathcal{F}_i=0.0$. System size is $N=201$.}
\label{fig2}
\end{figure}
 \begin{figure}[t]
 \includegraphics[width=0.98\columnwidth]{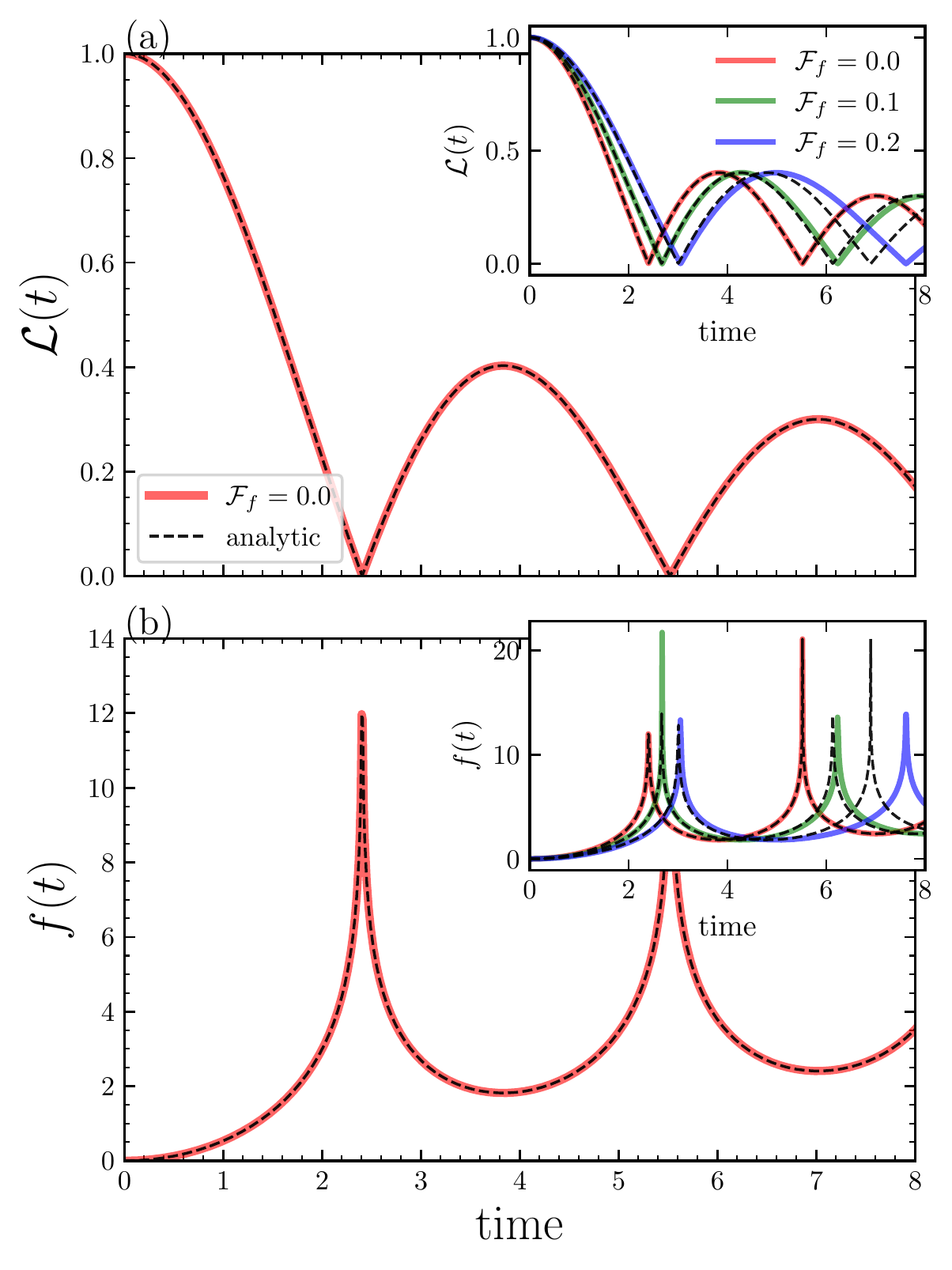}
 \caption{(colour online) Results for the backward quench from a localized state ($\mathcal{F}\neq0$) to a plane wave state ($\mathcal{F}=0$). (a) evolution of Loschmidt echo for the XX model with $\mathcal{F}_f=0.0$. (b) corresponding dynamical free energy. The black dashed curve corresponds to the analytical result $\mathcal{G}(t)=|J_0(Jt)|$. The initial state is fixed to be one of the localized states in the middle of the spectrum of the initial Hamiltonian with $\mathcal{F}_i=1.0$. System size is $N=201$. Insets show results for a type quench that does not ramp over the critical point, by setting the final field strength $\mathcal{F}_f\ne0$. Black dash lines are $J_0(|\mathcal{F}_f-\mathcal{F}_i|t)$. }
\label{fig3}
\end{figure}
\subsection{Non-interacting XX model}
Here we study the XX model which is subjected to a tilted magnetic field that varies linearly in position and changes sign at the center of the spin chain so that the ground state igiven by domain wall configuration. This model can be solved exactly by using the Jordan-Wigner transformation\cite{Elliott1961,Barouch1970,Barouch1971}. Rewriting the local spin-lowering and spin-raising operators $S_n^\pm=S_n^x\pm iS_n^y$, then with the Jordan-Wigner transformation $S_n^-=\prod_{m<n}(-1)^{c_m^\dagger c_m}c_n=\prod_{m<n}(1-2c_m^\dagger c_m)c_n$, Eq.(\ref{eq3}) with $J_z=0$ maps to the Wannier-Stark problem\cite{Case1973,Eisler2009,Lancaster2010} in a lattice subject to a constant force (or electric field) $\mathcal{F}$
\begin{equation}
H_{\rm xx} = -\frac{J}{2}\sum_
n\Big[c_n^\dagger c_{n+1}+c_{n+1}^\dagger c_n\Big]+\sum_n\mathcal{F}nc_n^\dagger c_n
\label{eq4}
\end{equation}
with introducing the following basis change
\begin{eqnarray}
\eta_m^\dagger&=&\sum_nJ_{n-m}(x)c_n^\dagger\nonumber\\
			&=&\frac{1}{\sqrt{N}}\sum_k\exp\Big[-ikm-ix\sin{(k)}\Big]c_k^\dagger
\label{eq5}
\end{eqnarray}
where $c_n=\frac{1}{\sqrt{N}}\sum_kc_ke^{ikn}$ , $N$ is the total number of sites and $J_n(x\equiv\frac{J}{\mathcal{F}})$ is a Bessel function of the first kind. One can arrive at the diagonalized Hamiltonian $H_{\rm xx}=\sum_m\epsilon_m\eta_m^\dagger\eta_m$ with $\epsilon_m=m\mathcal{F}, (m=0,\pm1,\dots)$.As $|J_n(x)|<e^{-|n|}$ for $x\ll n$, all of the eigenstates are (Bloch) localized for $\mathcal{F}\ne0$. Each eigenstate, $\eta_m^\dagger\ket{0}$ is localized close to the site $m$ with an inverse localization length given by $\zeta^{-1}\approx 2\sinh^{-1}(1/x)$. For the Bloch localization, there is no mobility edge in the spectrum and all eigenstates become localized, so the localization length is an energy-independent quantity. This is in contrast to the Anderson localization model, where the localization length is energy-dependent (smaller near the middle of the energy band) 
\par
Fig.\ref{fig1}-(a) show density profile $|\langle m|\Phi_n\rangle|^2$ and ground state particles occupation $\langle c_n^\dagger c_n\rangle$ as a function of the site index $n$. Results were extracted from different field strengths $\mathcal{F}$ numerically on a finite chain with length $N=201$, and we chose a state in the middle of the spectrum with $\epsilon_{m=N/2}=0$. With changing field from zero to some finite values, the density profile becomes localized about site in the middle of chain $i=N/2$ with localization length $\zeta^{-1}\approx 2\sinh^{-1}(\mathcal{F}/J)$. This corresponds to a delocalization-localization transition, such that particles freely located everywhere in the lattice for $\mathcal{F}=0$ become finally localized about site $n=N/2$. Interestingly, one can see from Fig.\ref{fig1}-(b), as the particles' occupation number evolves from extended at $\mathcal{F}=0$ to a perfect domain-wall profile at $\mathcal{F}=0.5$. The domain width is in great agreement with the localization length. 
\par
We study the nonequilibrium dynamics that arise when the field is suddenly tunned as a global quench in the model. By preparing the system as a state with $\epsilon_{m=N/2}=0$ of the Hamiltonian $H_{\rm xx}(\mathcal{F}_i=0)$ and then performing a sudden quench to the final Hamiltonian $H_{\rm xx}(\mathcal{F}_f)$, we examine the behavior of dynamical free energy. If $\mathcal{F}_i=0$, the system is described with plane wave states $\ket{\Psi_i(k)}\equiv\ket{k}=\frac{1}{\sqrt{N}}\sum_ne^{ikn}c_n^\dagger\ket{0}$, with wave vector in first Brillioum zone $k=\in(\pi,\pi]$, and eigenvalues $\epsilon_k=-J\cos{k}$. So the return amplitude can be read as 
\begin{eqnarray}
\mathcal{G}(t)&=&\bra{\Psi_i(k)}e^{-iH_{\rm xx}(\mathcal{F}_f)t}\ket{\Psi_i(k)}\nonumber\\
	&=&\sum_m\bra{k}e^{-iH_{\rm xx}(\mathcal{F}_f)t}\ket{m}\bra{m}\ket{k}\\
	&=&\sum_me^{-im\mathcal{F}_ft}|\bra{m}\ket{k}|^2=\frac{1}{N}\sum_me^{-im\mathcal{F}_ft}\nonumber
\label{eq6}
\end{eqnarray}
which  the following formula  used
\begin{eqnarray}
\bra{k}\ket{m}=&&\bra{0}c_k\eta_m^\dagger\ket{0}\nonumber\\
=&&\frac{1}{\sqrt{N}}\sum_{k'}e^{-ik'm-ix\sin{k'}} \underbrace{\bra{0}c_kc_{k'}^\dagger\ket{0}}_{\delta_{k,k'}}\nonumber\\
=&&\frac{1}{\sqrt{N}}e^{-ikm-ix\sin{k}}
\label{eq7}
\end{eqnarray}

where $\ket{m}$ is $m$-th eigenstate of final Hamiltonian which is localized state $\ket{m}=\sum_n\delta_{mn}\eta_n^\dagger\ket{0}$ with eigenvalue $\epsilon_m=m\mathcal{F}_f$.  With a little algebra one could show $\mathcal{L}(t)=|\sin{x}/x|^2$ ($x\equiv\mathcal{F}_ftN$), with a series of zeros at $t^*_n=\frac{2\pi n}{\mathcal{F}_fN}$ which indicates LE reaches zeros. 
\par
To validate the analytical results, we numerically explore the Loschmidt echo and the dynamical free energy. We set the initial state is set to be the ground state of $H(\mathcal{F}=0)$, and then a finite value $\mathcal{F}$ is switched on at $t = 0$. We first focus on the occurrence of zeros in the Loschmidt echo, which can be recognized as the signatures of DQPTs. It can see in Fig.\ref{fig2}-(a), that the Loschmidt echo becomes zero at some critical times, with the values agreeing well with the analytic prediction $t^*_n=\frac{2\pi n}{\mathcal{F}_fN}$. This demonstrates that Eq.(\ref{eq6}) is a good approximation.
\par
To show the zeros of $\mathcal{L}(t)$ more precisely, we also plot the dynamical free energy $f(t)$, which diverges at the dynamical critical time. As depicted in Fig.\ref{fig2}-(b), the numerical and analytical results are in well agreement, as $f(t)$ exhibits clear peaks at $t=t^*$.
\par
As shown in Fig.\ref{fig2}, we explored different field values $\mathcal{F}$. Before a quench ($\mathcal{F}_i=0$) particle states were defined by a plane wave and distributed over the chain. The model believes to be in a localized state for any finite field value $\mathcal{F}_f$. These quench clearly show DQPTs with ramping a quantum critical point. It should be mentioned that we also observe (not shown) DQPTs for the forward quench not ramping over the critical point, with shifting the critical times $t^*$.
 \begin{figure}[t]
 \includegraphics[width=1\columnwidth]{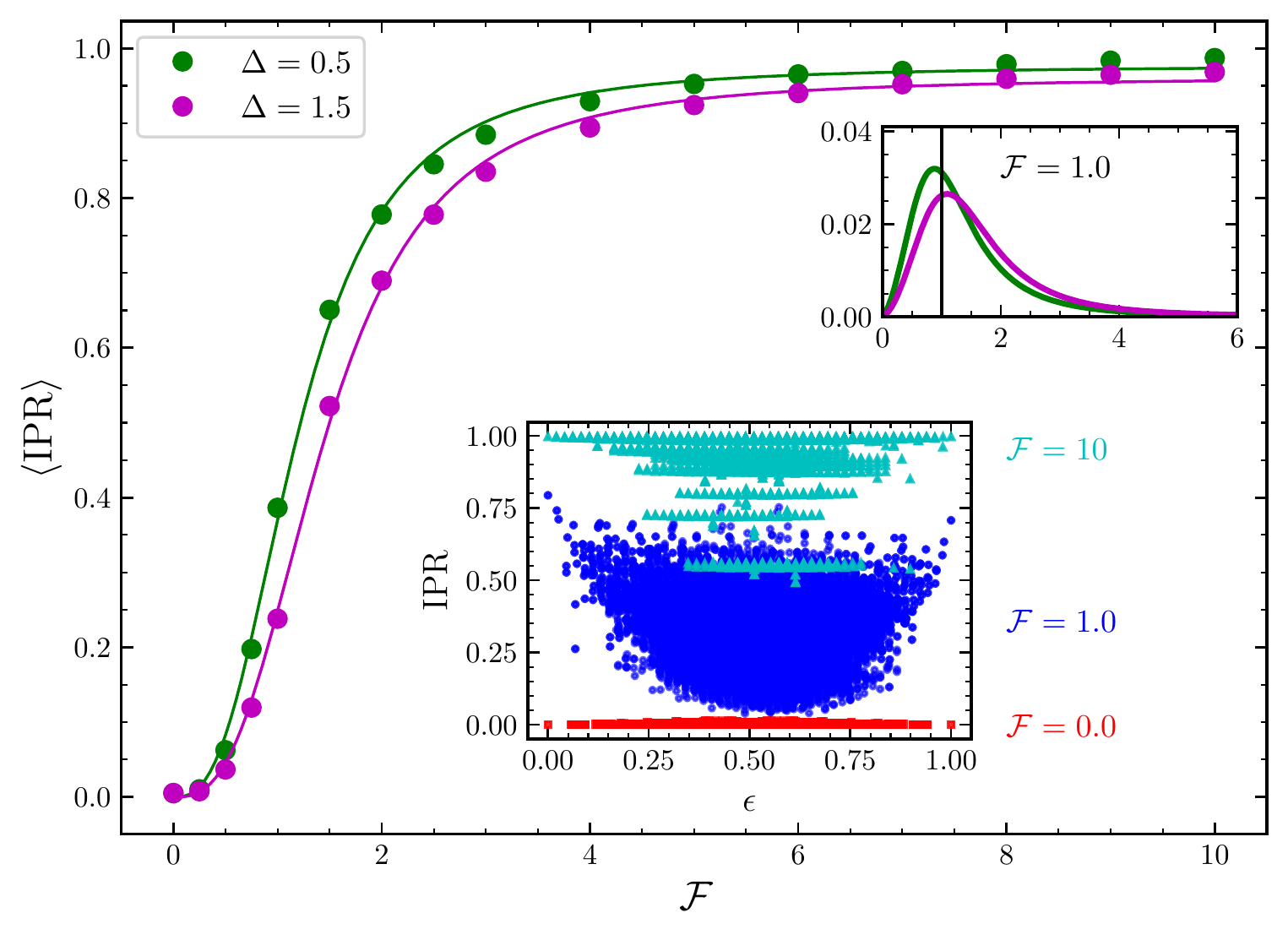}
 \caption{(colour online) Main panel shows the inverse participation ratio of $H_{\rm xxz}$ (Eq.~(\ref{eq8})) model averaged over all states versus magnetic field strength $\mathcal{F}$ for two exchange anisotropy parameters $\Delta=0.5, {\rm and}~1.5$. The solid lines are fitted curves and their derivatives are plotted in the top right inset panel. The bottom inset panel shows IPR for three different field values with $\Delta=0.5$. Data was obtained using the exact full diagonalization technique for chain length $N=14$.}
\label{fig4}
\end{figure}
\par 
Now, we let the model do a second quench in a reverse path from $H_{\rm xx}(\mathcal{F}_i\rightarrow1)$ to $H_{\rm xx}(\mathcal{F}_f\rightarrow0)$. Model ground state is a domain-wall and can be constructed by including all negative energy states $\ket{\Phi_i}=\prod_{m<0}\eta_m^\dagger\ket{0}$. Fig.\ref{fig1}-(b) shows the density profiles $\langle c_n^\dagger c_n\rangle=\sum_{m<0}|J_{n-m}(x)|^2$ in the ground state for different characteristic length $x\equiv\frac{J}{\mathcal{F}}$ (see details in appendix~\ref{appB}). But as the case of forward quench and for simplicity, we consider $m$-th eigenstate of final Hamiltonian which is localized state $\ket{m}=\sum_n\delta_{mn}\eta_n^\dagger\ket{0}$. Having this, the return amplitude can be formulated as 
\begin{eqnarray}
\mathcal{G}(t)&=&\bra{m}e^{-iH_{\rm xx}(\mathcal{F}_f=0)t}\ket{m}\nonumber\\
	&=&\sum_k\bra{m}e^{-iH_{\rm xx}(\mathcal{F}_f=0)t}\ket{k}\bra{k}\ket{m}\nonumber\\
	&=&\sum_k e^{-iJt\cos{k}} |\bra{k}\ket{m}|^2
\label{eq8}
\end{eqnarray}
with wave vector in first Brillioum zone $k=\in(\pi,\pi]$, we can  replace  the summation by the integration
\begin{equation}
\mathcal{G}(t)=\frac{1}{2\pi}\int_{-\pi}^\pi e^{-iJt\cos{k}}dk=J_0(Jt)
\label{eq9}
\end{equation}
where $J_0(Jt)$ is the zero-order Bessel function, and has series of zeros $x_n$ with $n=0,1,\cdots$. This indicates the Loschmidt echo reaches zeros at times $t^*_n=x_n/J$, which is independent of field strength $\mathcal{F}$. This is similar to the quench addressed in the lieatrature\cite{Yang2017,Hao2019}.
\par
In the main panel of Fig.\ref{fig3}, we present numerical and analytical results of the Loschmidt echo and the dynamical free energy for a quench ramped over the critical point. The initial state sets to be an eigenstate of XX model with $\epsilon_{m=N/2}=0$ which $\mathcal{F}_f=1.0$, and a quench to a final state with $\mathcal{F}_f=0.0$ is applied. The numerical and analytical predictions are in agreement and DQPTs happen at a series of times related to the Bessel function's zeros. Moreover, results for a type of quench that does not ramp over the quantum critical point are shown in the insets. Interestingly, we found the Loschmidt echo becomes zero at some critical times, and correspondingly the dynamical free energy $f(t)$ diverges at the dynamical critical time. It can be seen critical time $t^*$ shifts to bigger time and we found numerically that $t^*\sim x_n/|\mathcal{F}_f-\mathcal{F}_i|J$, which is a good approximation for short times. 
\subsection{Interacting case $J_z\neq0$}
To address the interaction (Ising term) effect on the Loschmidt echo dynamic,  we consider  the XXZ model:
\begin{equation}
H_{\rm xxz} =J\sum_
n\Big[S_n^xS_{n+1}^x+S_n^yS_{n+1}^y+\Delta S_n^zS_{n+1}^z+h_nS_n^z\Big]
\label{eq8}
\end{equation}
here $J$ is the exchange coupling constant, and $J_z=\Delta$ parametrizes the exchange anisotropy. For zero magnetic field $h=0$, the model is solvable by Bethe ansatz\cite{Bethe1931}. It has an infinite number of conservation laws, and affect the thermalization of the quantum system for longer times after the quantum quench. The phase diagram consists of a gapped ferromagnetic (FM) for $\Delta<-1$, a gapless Luttinger liquid (LL) for $-1<\Delta<1$, and antiferromagnetic (AFM) for $\Delta>1$ phase\cite{takahashi_1999}. 
\par
Previous analytical and numerical results on interacting many-body fermions and spins-$1/2$ \cite{Enss2012,Andraschko2014} show that there are no dynamical quantum phase transitions for quenches inside the LL phase and the AFM phase, while there exist oscillations of the return amplitude with time. On the other hand, the quantum quenches across the transitions between the LL and the gapped FM and AFM phases (including the Berezinsky-Kosterlitz-Thouless (BKT) point) show dynamical quantum phase transitions, according to the numerical calculaton\cite{Andraschko2014}. With this may, one may conclude that interactions in integrable one-dimensional spin models do not destroy dynamical quantum phase transitions. So it motivates us to probe DQPT in the Stark ladder chain in the presence of interactions.
\par
 In the following, we consider $\Delta>0$ and quench the model concerning the linear magnetic field. We also investigate the possible effect of the BKT transition at $\Delta=1$ on DQPT. To treat the many-body problem and consider the interacting term, we choose the time-dependent density matrix renormalization group (tDMRG) technique implemented using the C++ iTensor library\cite{itensor}. We approximate the evolution operator by a matrix-product operator (MPO)\cite{Zaletel2015} with a $4$-th order\cite{Bidzhiev2017} Trotter scheme.
 \begin{figure}[t]
 \includegraphics[width=1\columnwidth]{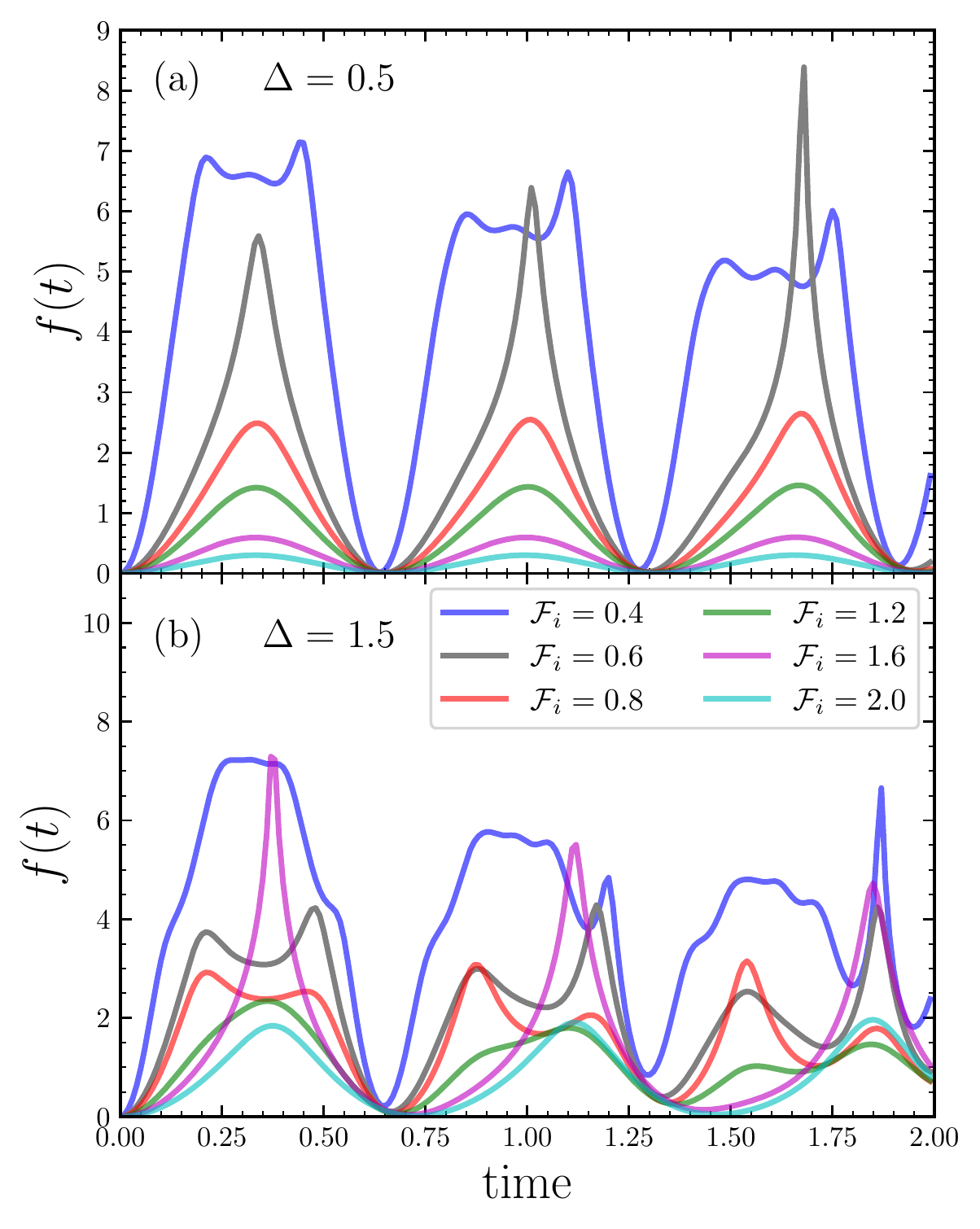}
 \caption{(colour online) Evolution of dynamical free energy for the XXZ model with (a) $\Delta=0.5$ and (b) $\Delta=1.5$ in forward quench. The initial state sets as the ground state of the Hamiltonian with different field values $\mathcal{F}_i=0.4,\dots,2.0$ and the model quenched to a final Hamiltonian with a field value $\mathcal{F}_f=10$. The system size is $N=101$.}
\label{fig5}
\end{figure}
 \begin{figure}[t]
 \includegraphics[width=1\columnwidth]{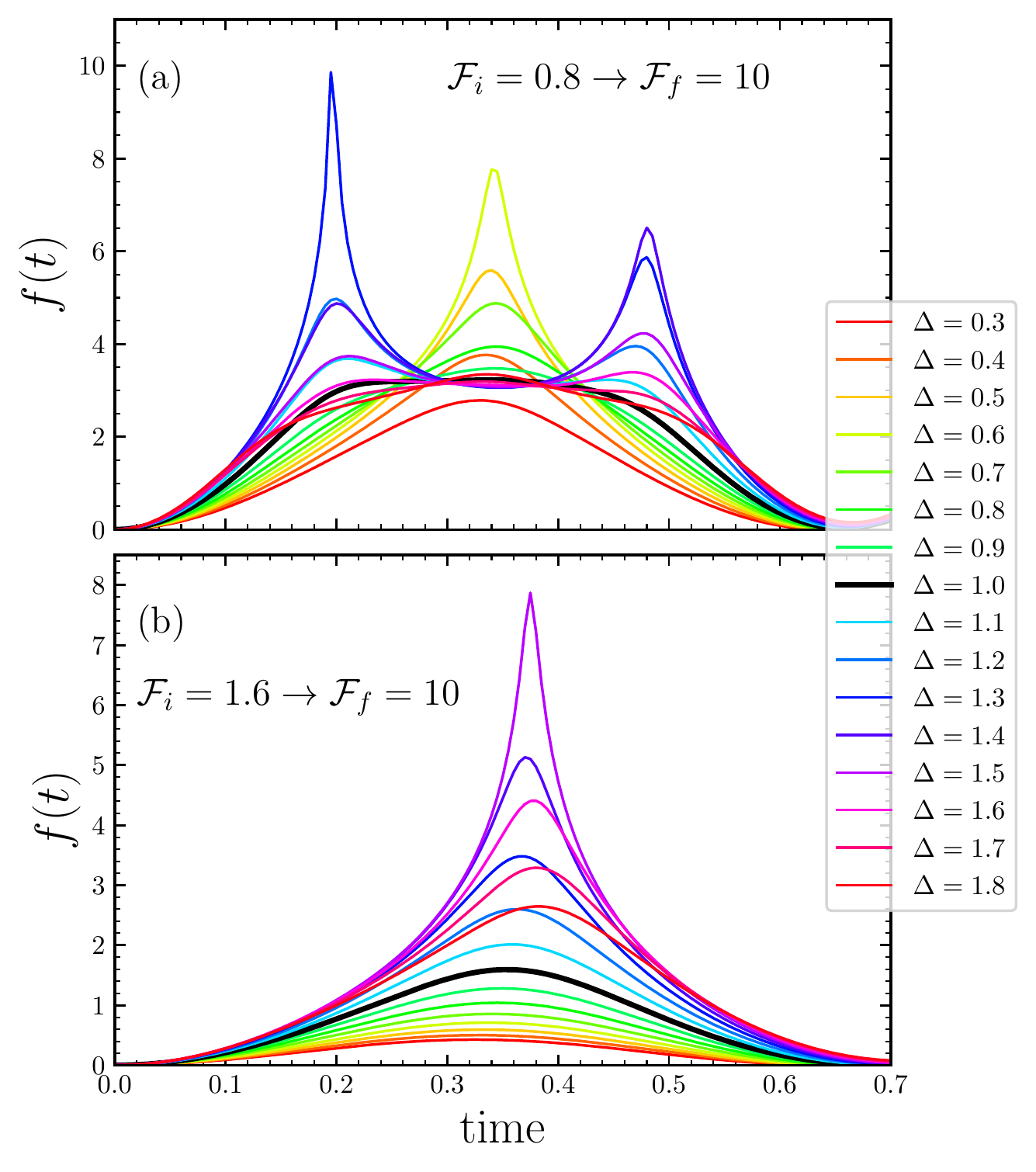}\\
 \caption{(colour online)Dynamical free energy for XXZ model for different anisotropic parameter $\Delta=0.3,\dots,1.8$. The initial state sets as the ground state of the Hamiltonian with (a) $\mathcal{F}_i=0.8$ (b) $\mathcal{F}_i=1.6$, and the final Hamiltonian fixed with $\mathcal{F}_i=10$. The system size is $N=101$. Numerical results obtined by excat diagonalization are presented in appendix~\ref{appA}. }
\label{fig6}
\end{figure}
 \begin{figure}[t]
 \includegraphics[width=1\columnwidth]{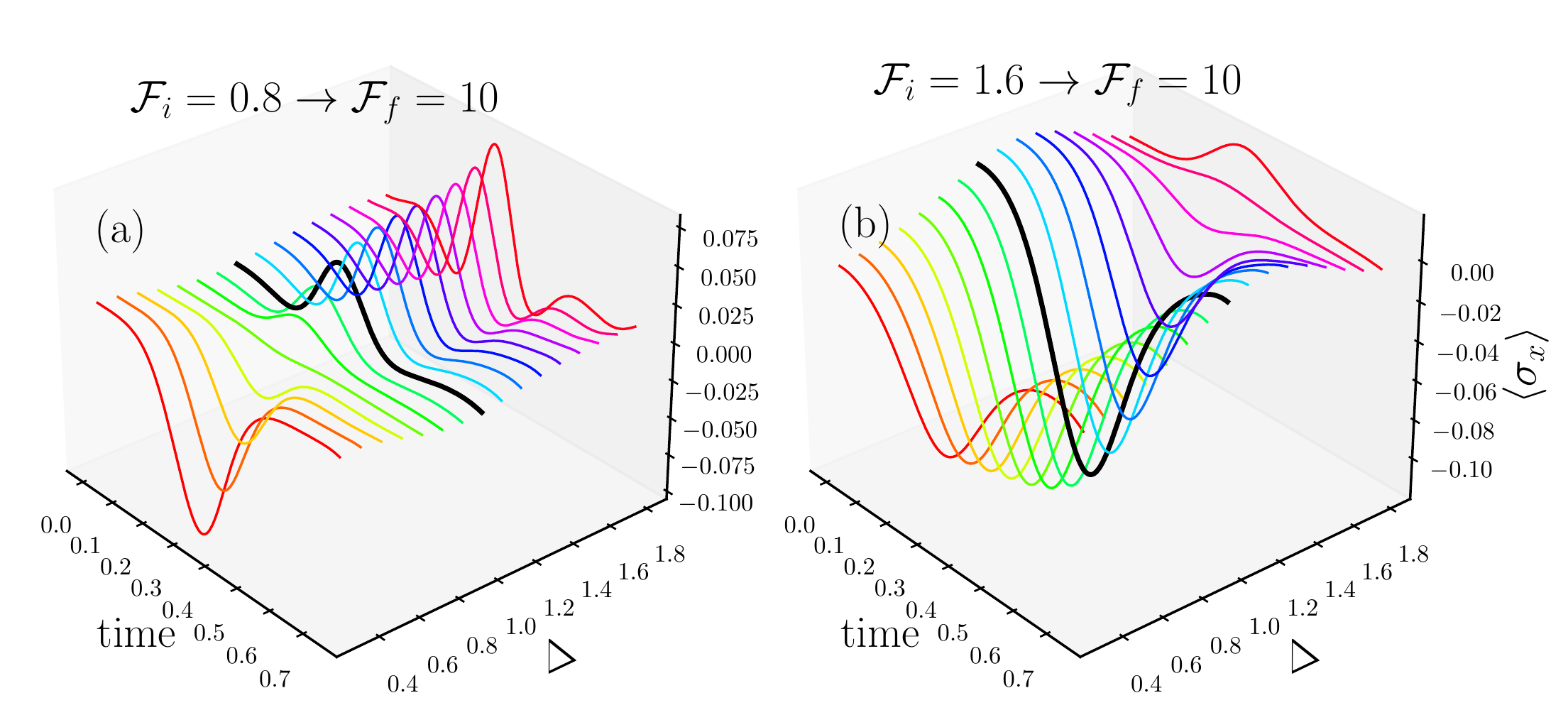}\\
  \includegraphics[width=1\columnwidth]{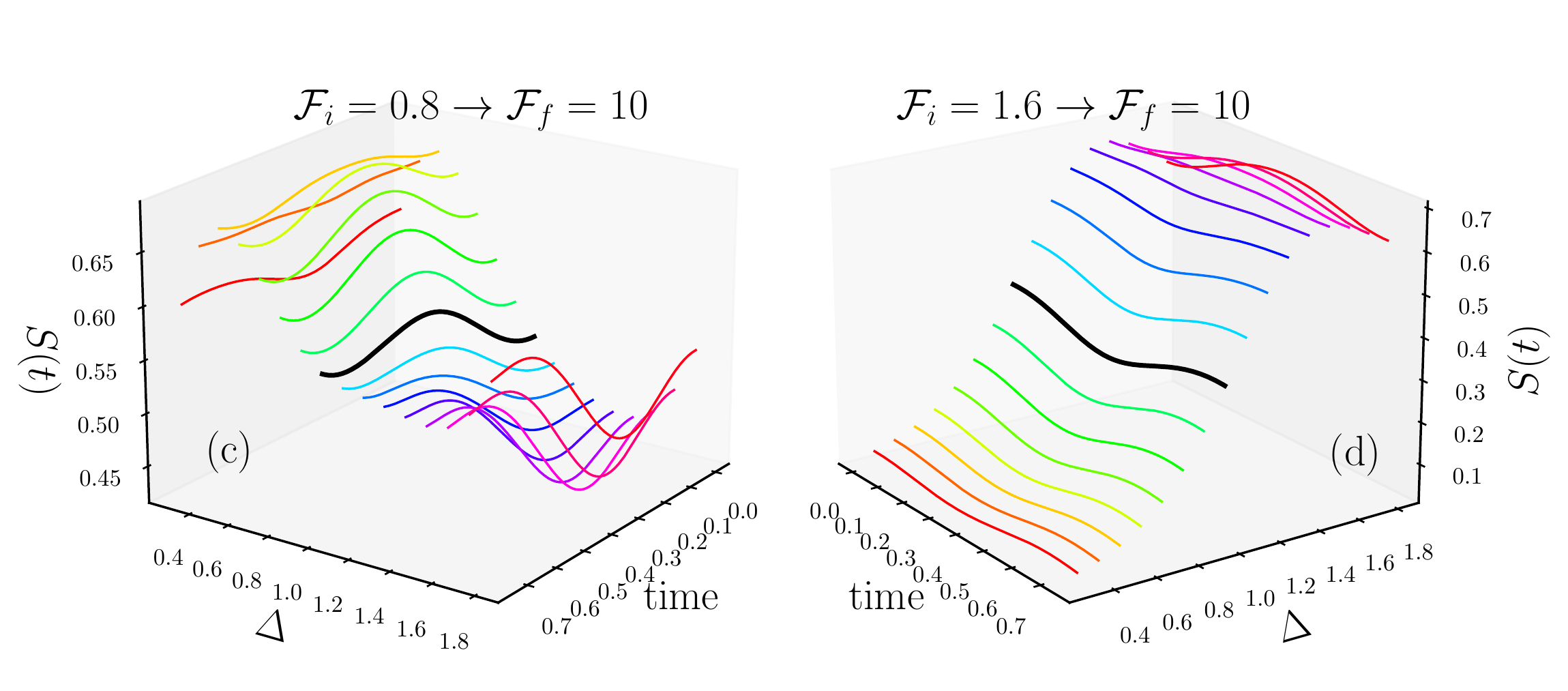}
 \caption{(colour online) (a) and (b) local magnetization along $x$ as a function of time. (c) and (d) half-chain entanglement entropy  evolution.  Parameters set as in Fig.\ref{fig6}.}
\label{fig7}
\end{figure}
 \begin{figure*}[t]
   \includegraphics[width=2\columnwidth]{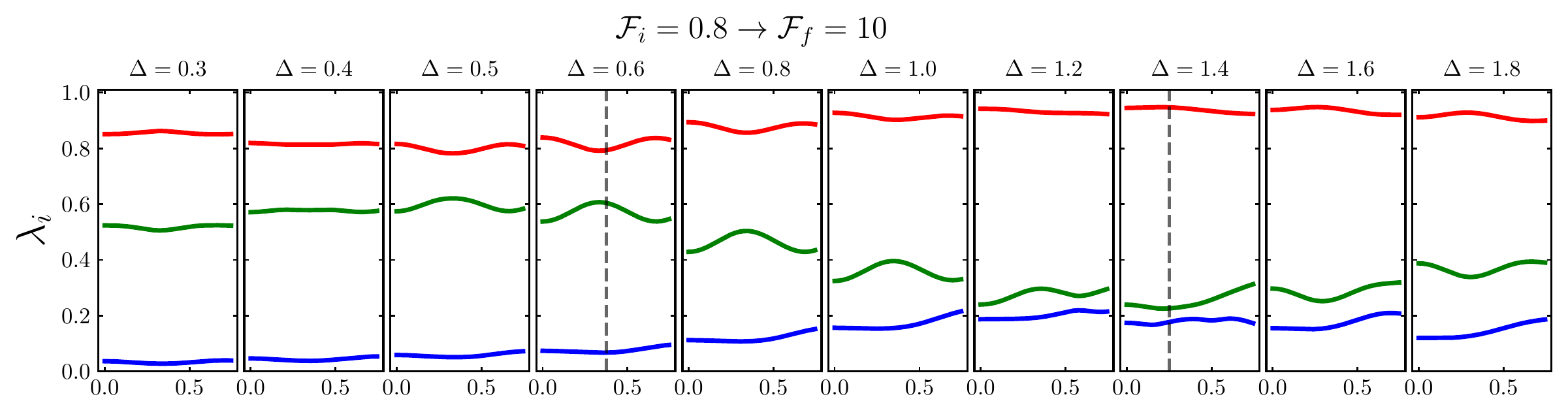}\\
  \includegraphics[width=2\columnwidth]{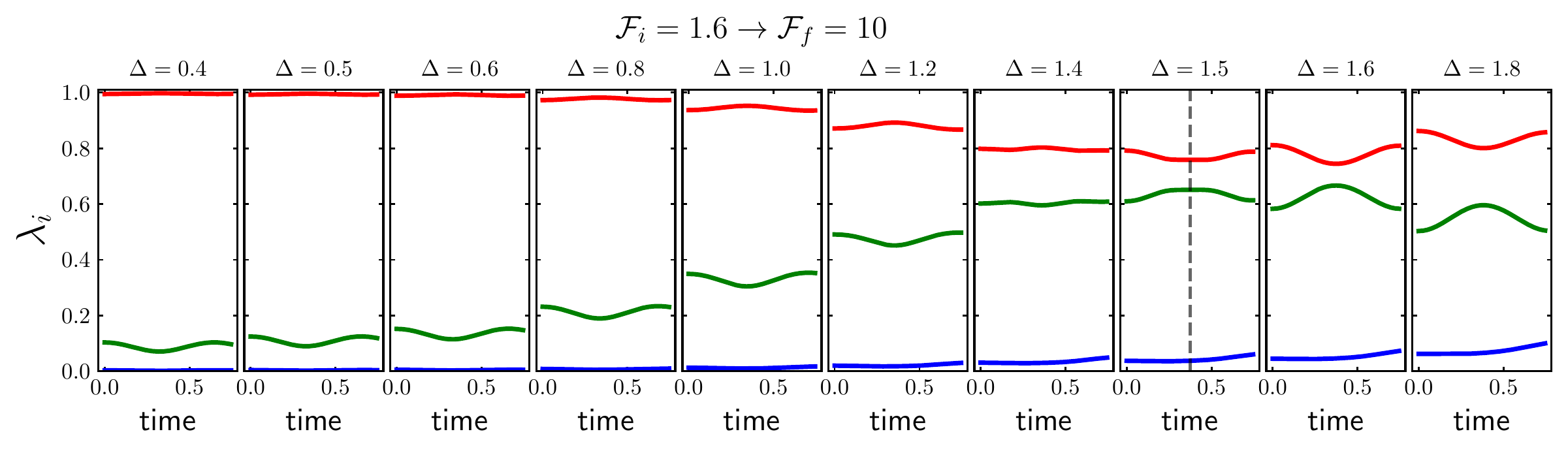}
 \caption{(colour online) Evolution of the entanglement spectrum (three highest singular
values of the Schmidt decomposition across a bond at the center of the chain). Top and bottom panels are correspond to the quenches with $\mathcal{F}_i=0.8\rightarrow\mathcal{F}_f=10$ and $\mathcal{F}_i=1.6\rightarrow\mathcal{F}_f=10$, respectively. Parameters set as in Fig.\ref{fig6}.}
\label{fig8}
\end{figure*}
\par
Before exploring the dynamical phase of the model, it would be intuitive to have a look at the Bloch localization-delocalization transition in the interacting case, as usually interaction explodes accessible states in the model and tends to delocalize the system. However, it has been recently predicted an ergodic to many-body localization transition occurs in this model, even a very small value of disorder strength in this model ~\cite{Refael2019}. Motivated by this, we explore the inverse participation ratio (IPR) which is a measure of localization-delocalization of a quantum mechanical system~\cite{Misguich2016}. The IPRs have been used to measure the localization in the context of the Anderson localization \cite{Mirlin2008}, and have also been useful to detect the MBL\cite{Luca2013}. A basic definition of IPR can be given as follows
\begin{equation}
{\rm IPR}=\sum_{k}^{\mathcal{D}}{\rm IPR}_k=\sum_{i,k}^{\mathcal{\mathcal{D}}}|\langle a_i|\psi_k\rangle|^4
\label{eq9}
\end{equation}
The IPR quantifies how much the eigenstates are localized in the preferential basis $|a_i\rangle_{i=1,\dots,\mathcal{D}}$. They can range from ${\rm IPR}_{\rm max}=1$ (completely localized state) to ${\rm IPR}_{\rm min}=1/\mathcal{D}$ (maximally delocalized limit). Fig.\ref{fig4} shows how the IPR (as averaged over all states) changes with magnetic field strength $\mathcal{F}$. It is apparent that the model is delocalized $\langle {\rm IPR}\rangle\rightarrow0$ with $\mathcal{F}/J\rightarrow0$ and localized $\langle {\rm IPR}\rangle\rightarrow1$ for $\mathcal{F}/J\rightarrow\infty$. Determining the critical point needs a proper finite size scaling, which is beyond the goal of the present paper. However, we anticipate a quantum critical point located $\mathcal{F}_c(\Delta)/J\le1.0$, which the model undergoes a localization-delocalization transition (see the top right inset in Fig.\ref{fig4}). To justify our prediction, we refer the reader to Ref.[\cite{Refael2019}] in which for a week random on-site potential a many-body localization derived by $\mathcal{F}_c(\Delta=1)/J\approx1.0$ is reported. 
\par
As stated before, we set the model parameters to quench with and without ramping through a quantum critical point. In Fig.\ref{fig5}, we numerically present the dynamical free energy for two exchange anisotropy parameters $\Delta=0.5$ and $\Delta=1.5$. The initial magnetic field changes in such a way as the model exists in the delocalized ground state for $\mathcal{F}_i=0.4$ and localized state for $\mathcal{F}_i=2.0$ (but still close to the critical point $\mathcal{F}_c\approx1.0$). Then we quench the model to the deep localized phase with a final magnetic field $\mathcal{F}_f=10$. In all quenches, in this part, we only consideFig.r the ground state of the model as an initial state. 
\par
The details of how the Loschmidt echo and the dynamical free energy evolves in time depending on whether the system is in the gapless LL phase ($\Delta<1.0$) or the gapped AF phase ($\Delta>1.0$). We notice that when the model only exists in the gapless phase, with a quench ramping through the quantum critical point $\mathcal{F}_c\approx1.0$, the dynamical free energy $f(t)$ diverges at the dynamical critical times (see Fig.\ref{fig5}-(a)). However, as the numerical results display when the model exists in the gapped phase (within the set of parameters chosen here), both quench with and without ramping through the quantum critical point signal the occurrence of DQPTs. The dynamical free energy $f(t)$ shows divergence at the dynamical critical times, as depicted in Fig.\ref{fig5}-(b), for magnetic field value $\mathcal{F}_i=0.6$ ($\mathcal{F}_i=1.6$), which is located in the delocalized (localized) phase.
\par
We also explore the effects of interaction  on DQPTs by tunning the exchange anisotropy parameter $\Delta$. As it is shown in Fig.\ref{fig6}, we do quench with $\mathcal{F}_i=0.8~{\rm or}~1.6\rightarrow\mathcal{F}_f=10$. The distinct effect of interaction is clear when $\Delta$ increases gradually.  We noticed two different behaviors depending on the quench with or without ramping through the localization-delocalization critical point. Indeed, for a quench in the same localized phase Fig.\ref{fig6}-(b), by increasing $\Delta$ we recognized an amplifying of divergence in the dynamical free energy $f(t)$ with a tiny shift in  dynamical critical time (let us call $t^*_0$) to later times. The biggest nonanalyticity happens at $\Delta\simeq1.5$, and then the dynamical free energy fades. However, with a quench through the delocalized-localized point, see Fig.\ref{fig6}-(a), we identify two DQPTs for $\Delta\simeq0.6<1.0$ and $\Delta\simeq1.3>1.0$ correspond to the divergency of dynamical free energy. Intrestingly, we find that dynamical critical time $t^*$ happens sooner for $\Delta>1.0$.  
\par
The realized properties sound interesting and worth more exploration. To do so, we came up with the idea to find a connection to equilibrium concepts, such as order parameters, or establish a connection to entanglement production in the model. Indeed, the relation of DQPTs with equilibrium properties, such as local observables was addressed\cite{Trapin2018}, however, the general connection between DQPTs and local expectation values remains ambiguous. Connections to the entanglement entropy have also been studied. It is argued that DQPTs may relate to the rapid growth\cite{Jurcevic2017} or peaks\cite{Schmitt2018} in the entanglement entropy. In some integrable models, a connection between DQPTs and crossings in the entanglement spectrum is also recognized\cite{Elena2014}.
\par
Fig.\ref{fig7} (a) and (b) present time evolution of the average local magnetization $\langle\sigma_x\rangle$, for both quenches done in Fig.\ref{fig6}. The general tendency of $\langle\sigma_x\rangle$ over the time window shown here reveals a peak about the critical time corresponds to the non-analicity of dynamical free enrergy. Interestingly, we found for $\Delta\simeq0.8$ (in Fig.\ref{fig7} (a) ) and $\Delta\simeq1.5$ (in Fig.\ref{fig7} (b) ), $\langle\sigma_x\rangle$ changes polarization direction corresponds to the strongest divergence in $f(t)$, note that this change is time independent and drives via the parameter $\Delta$, let us call it type-I for later references. However, for the quench ramped over critical field (namely $\mathcal{F}_i=0.8\rightarrow\mathcal{F}_f=10$), we noticed the strong divergence in $f(t)$ corresponds to a sudden changes in the $\langle\sigma_x\rangle$ polarization with time, we call it type-II. 
\par
 Now, let us explore the entanglement entropy and possible connection to the above observations. The bipartite entanglement entropy, $S=-\sum_ip_i\log{p_i}$, where $p_i=\lambda_i^2$s are singular values of the Schmidt decomposition across a bond, is easily accessed through the DMRG calculation. Numerical results present in Fig.\ref{fig7} are the half-chain entanglement entropy corresponding to the quench experiment done in Fig.\ref{fig6}. We find, that depending on whether the quench happens through the critical point or not, different entanglement production and evolution in the model. For quench in the same localized phase, see Fig.\ref{fig7}-(b), it is clear that maximum entanglement entropy produced for $\Delta\simeq1.5$ where we recognized the nonanalyticity in $f(t)$. The situation for quench amid the delocalized-localized point is a bit different. While for a quench with $\Delta<1.0$, we find the nonanalyticity in $f(t)$ corresponds peak in the entanglement production (within the parameters set and time window shown here), in the case with $\Delta>1.0$ we do not observe such correspondence. However, sudden changes in the entanglement entropy are seen in the vicinity of the critical time.  
\par
In spin chains, the evidence of DQPTs has been connected to the vanishing Schmidt gaps\cite{Elena2014}, and very recently two DQPTs were distinguished based on the Schmidt gaps in the entanglement spectrum\cite{Nicola2021}. So to make our study and observation more concrete, it also worths having a look at the entanglement spectrum $\lambda_i$, as the data is already available.  In Fig.\ref{fig7} results are drawn for the three biggest Schmidt values. It is readily apparent,  in the evolution of the Schmidt gaps,  that the minimum gap occurs in the vicinity of the DQPTs. For instance for the first Schmidt gap, defined as $\lambda_1-\lambda_2$, we noticed a connection with the peak occurs in the entanglement entropy and the type-I change of the local magnetization, for $\Delta\simeq0.6$ ($\Delta\simeq1.5$) in quench with $\mathcal{F}_i=0.8\rightarrow\mathcal{F}_f=10$ ($\mathcal{F}_i=1.6\rightarrow\mathcal{F}_f=10$). Likewise, for the case $\Delta>1.0$ in quench with $\mathcal{F}_i=0.8\rightarrow\mathcal{F}_f=10$, the second Schmidt gap, $\lambda_2-\lambda_3$, develops minimum close to the DQPTs.\\

\section{Summary and Outlook}\label{sec3}
The well-known Wannier-Stark effect, related to the localization of the noninteracting particles in the presence of a linear potential, has recently received attention in the context of interacting spins and fermions, and a new type of localization known as Stark many-body localization is introduced. Here, we consider spin chains in the presence of a linear potential, which induces delocalization-localization transition even in one dimension. We study dynamical quantum phase transition (DQPT) due to sudden global quenches across a quantum critical point when the system undergoes a localization-delocalization transition. To do this, we consider the XX and XXZ spin chains, and explore two type quenches with and without ramping through the delocalization-localization critical point. The XX model was mapped to the free fermion particles, so both analytical and numerical results of quench were provided. Results unveil that the dynamical signature of localization-delocalization transition can be characterized by the nonanalyticities appear in dynamical free energy (corresponds to the zero points in the Loschmit echo). We also explore the interaction effects considering XXZ spin chains, using the time-dependent extension of the numerical DMRG technique. Our results reveal that depending on the anisotropic parameter $\Delta\lessgtr1.0$ (namely the Ising term), if both the initial and post-quench Hamiltonian are in the same phase or not, DQPTs may happen. We provide more analysis on the feature of DQPTs, in both the two type quenches, by connecting them to the average local magnetization, entanglement entropy production, and the Schmidt gap. We noticed two polarization direction changes (called them type-I and type-II ) in the local magnetization $\langle\sigma_x\rangle$ correspond to the nonanalyticities in dynamical free energy. Likewise, we observed peaks and sudden increases in the entanglement entropy accompanied by the decreasing the first and second gap evolution of the entanglement spectrum happens in the vicinity of DQPTs.

\section{Acknowledgement}
We thank Youcef Mohdeb and Saeed Mahdavifar for stimulating discussion and useful comments. J. V acknowledge support from Deutsche Forschungsgemeinschaft (DFG) KE-807/22-1. 

\begin{figure}[t]
 \includegraphics[width=1\columnwidth]{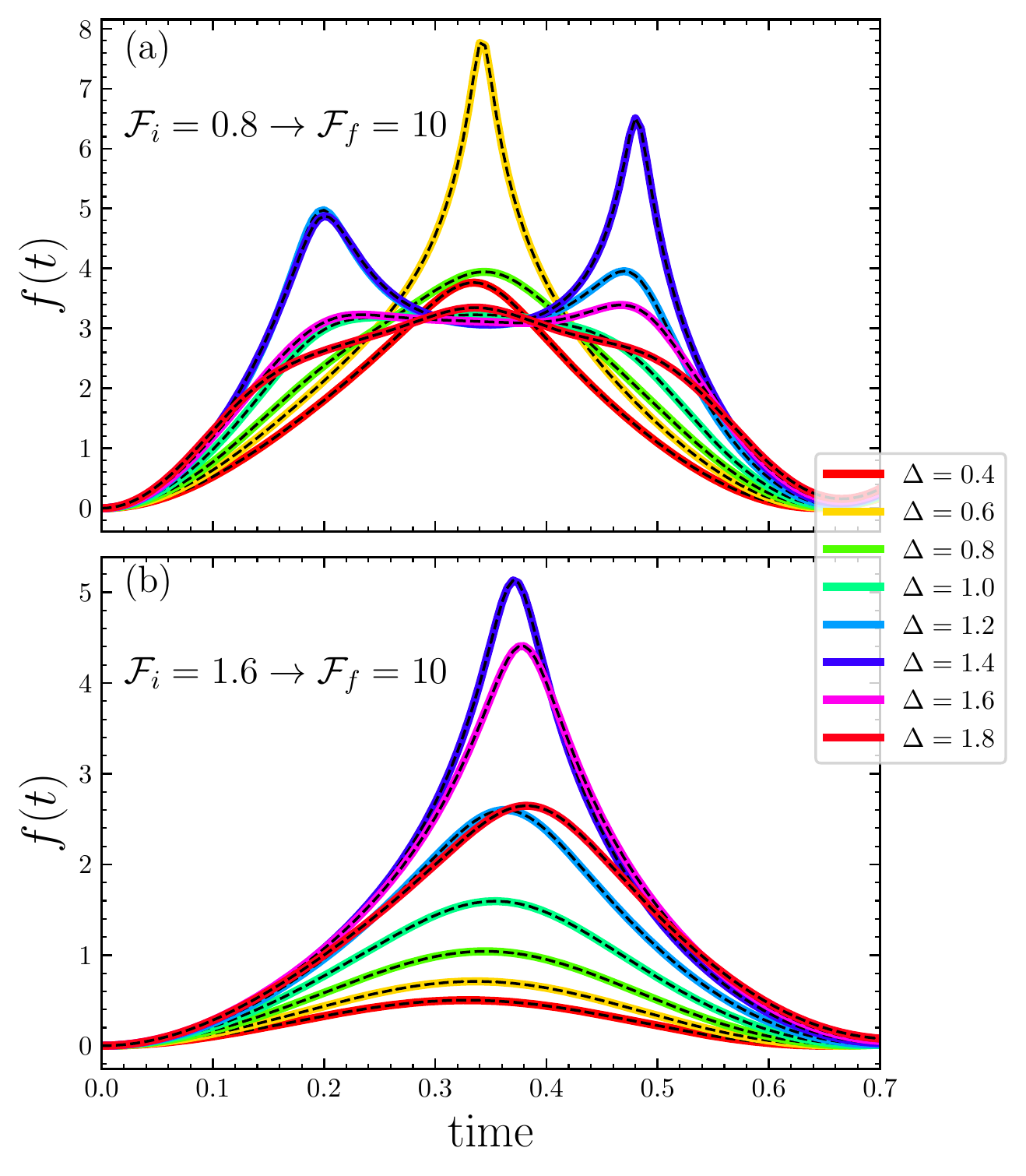}\\
 \caption{(colour online) Comparing results obtained numerically with the tDMRG (solid color lines) for system size $N=101$ and the data collected from numerical exact diagonalization method (black dashed lines) for system size $N=15$. Other parameters are set as the Fig.\ref{fig6}.}
\label{fig9}
\end{figure}
\appendix

\section{Finite size study}\label{appA}
Here we aim to show results presented for the XXZ case in the main text are not suffering the finite size effect. In Fig.\ref{fig9}, we plot the evolution of dynamical free energy computed numerically with tDMRG and exact diagonalization techniques. Interestingly, we notice a perfect agreement between the two techniques as the model is completely free of possible size effects.

\section{Details of the computation of  $\left<c^\dagger_nc_n\right>$}\label{appB}
In this appendix, we give details of the derivation of local observable Eq.(\ref{eq4}) in the main text. To do this, we start with the basis change definition given in Eq.(\ref{eq5}) as follows
\begin{eqnarray}
\eta^\dagger_m=\sum_nJ_{n-m}(x)c_n^\dagger\quad{\rm and }\quad\eta_m=\sum_nJ^*_{n-m}(x)c_n\nonumber\\
\end{eqnarray}
and having the ground state $\ket{\Phi_0}=\prod_{m<0}\eta_m^\dagger\ket{0}$, we are looking to find a closed formula for 
\begin{eqnarray}
\left<c^\dagger_nc_n\right>=\bra{\Phi_0}c^\dagger_nc_n\ket{\Phi_0}=\bra{0}\prod_{m'<0}\eta_{m'}~c^\dagger_nc_n~\prod_{m<0}\eta_m^\dagger\ket{0}\nonumber\\
\label{eq1a}
\end{eqnarray}
where $\{c_n,c_{n'}\}=\delta_{n,n'}$. We start by constructing the commutation between to basis

\begin{eqnarray}
c_n\eta_m^\dagger=&&c_n\sum_{n'}J_{n'-m}(x)c_{n'}^\dagger=\sum_{n'}J_{n'-m}(x)c_nc_{n'}^\dagger\nonumber\\
=&&\sum_{n'}J_{n'-m}(x)[\delta_{n,n'}-c_{n'}^\dagger c_n]=J_{n-m}(x)-\eta_m^\dagger c_n\nonumber\\
\label{eq2a}
\end{eqnarray}

note that $m=-L,-L+1,\cdots,L-1,L$ corespods to  system size  $2L+1$, then
\begin{widetext}
\begin{eqnarray}
c_n\prod_{m<0}\eta_m^\dagger=&&[J_{n+L}(x)-\eta_{-L}^\dagger c_n]\prod_{m>-L}^{<0}\eta_m^\dagger=J_{n+L}(x)\prod_{m>-L}^{<0}\eta_m^\dagger-\eta_{-L}^\dagger c_n\prod_{m>-L}^{m<0}\eta_m^\dagger\nonumber\\
=&&J_{n+L}(x)\prod_{m>-L}^{<0}\eta_m^\dagger-\eta_{-L}^\dagger[J_{n+L-1}(x)-\eta_{-L+1}^\dagger c_n]\prod_{m>-L+1}^{<0}\eta_m^\dagger\nonumber\\
=&&J_{n+L}(x)\prod_{m>-L}^{<0}\eta_m^\dagger-J_{n+L-1}(x)\prod_{m\ne-L+1}^{<0}\eta_m^\dagger+\eta_{-L}^\dagger\eta_{-L+1}^\dagger c_n\prod_{m>-L+2}^{<0}\eta_{-L+2}^\dagger\nonumber\\
=&&\dots\nonumber\\
=&&\sum_{l=0}^{L}(-1)^lJ_{n+l}(x)\prod_{m\ne l}^{<0}\eta_m^\dagger+(-1)^L\prod_{m<0}\eta_m^\dagger c_n
\label{eq3a}
\end{eqnarray}

we know $c_n\ket{0}=0$, so the second term can be ignored. 

\begin{eqnarray}
\left<c^\dagger_nc_n\right>=&&\bra{0}\prod_{m'<0}\eta_{m'}~c^\dagger_nc_n~\prod_{m<0}\eta_m^\dagger\ket{0}=\bra{0}  \left[ \sum_{l'=0}^{L}(-1)^{l'}J_{n+l'}^*(x)\prod_{m'\ne l'}^{<0}\eta_{m'}\right]~\left[ \sum_{l=0}^{L}(-1)^lJ_{n+l}(x)\prod_{m\ne l}^{<0}\eta_m^\dagger\right]\ket{0}\nonumber\\
\label{eq4a}
\end{eqnarray}
\end{widetext}

The Hamiltonian eingenvectors, namely Bessel function $J_{n-m}(x)$, are orthogonal and normalized,  $$\bra{0} \prod_{m'\ne l'}^{<0}\eta_{m'} \prod_{m\ne l}^{<0}\eta_m^\dagger\ket{0}=\delta_{l,l'}\delta_{m,m'}$$, so finally we arrive a compact formula as follows

\begin{eqnarray}
\left<c^\dagger_nc_n\right>=&&\sum_{l=0}\sum_{l'=0}(-1)^{l+l'}J_{n+l'}^*(x)J_{n+l}(x)\nonumber\\
=&&\sum_{l=0}J_{n+l}(x)^2
\label{eq5a}
\end{eqnarray}

Other useful formulas can also be extracted easily
\begin{eqnarray}
\left<c^\dagger_nc_m\right>=\sum_{l=0}J_{n+l}(x)J_{m+l}(x)\nonumber\\
\left<c_nc_m^\dagger\right>=\sum_{l=0}J_{n-l}(x)J_{m-l}(x)
\label{eq6a}
\end{eqnarray}

Using a sum rule, Eq.(\ref{eq6a}) can be rewritten as a simple product of Bessel functions. Dropping the arguments $x$ it reads
\begin{eqnarray}
\left<c^\dagger_nc_m\right>&=&\frac{x}{2(m-n)}\left[J_{m-1}(x)J_n(x)-J_m(x)J_{n-1}(x)\right]\nonumber\\
\left<c_nc_m^\dagger\right>&=&\frac{x}{2(m-n)}\left[J_{m+1}(x)J_n(x)-J_m(x)J_{n+1}(x)\right]\nonumber\\
\label{eq7a}
\end{eqnarray}

\bibliography{Ref/references}

\end{document}